\documentclass[conference]{IEEEtran}
\IEEEoverridecommandlockouts
\usepackage{cite}
\usepackage{amsmath,amssymb,amsfonts}
\usepackage{algorithm}
\usepackage{algpseudocode}
\usepackage{graphicx}
\usepackage{textcomp}
\usepackage{xcolor}
\usepackage{mathpartir}
\usepackage[T1]{fontenc}
\usepackage[scaled=0.8]{roboto-mono}

\usepackage{booktabs}          % \toprule, \midrule, \bottomrule
\usepackage{multirow}          % merged cells
\usepackage{threeparttable}    % table footnotes (optional but recommended)
\usepackage{pifont}
\usepackage{graphicx}          % \rotatebox (usually already loaded)
\usepackage{xspace}
\usepackage{balance}
\usepackage{hyperref}
\usepackage{enumitem}
\usepackage{subcaption}
\usepackage{framed}

\newcommand{\tool}{Shar\xspace}
\newcommand{\dsl}{Mystra\xspace}

\usepackage{ulem}

\definecolor{Azure}{RGB}{52, 152, 219}
\definecolor{Brick}{RGB}{169, 50, 38}
\definecolor{Pine}{RGB}{30, 132, 73}
\definecolor{FlowBlue}{RGB}{41, 128, 185}
\definecolor{CodeGreen}{HTML}{6A9955}
\newcommand{\crarrow}{\; \textcolor{Azure}{\rightarrow} \;}
\newcommand{\cDownarrow}{\;\; \textcolor{Brick}{\Downarrow} \;\;}
\newcommand{\cRarrow}{\, \textcolor{Pine}{\Rightarrow} \,}
\newcommand{\rulename}[1]{\textcolor{FlowBlue}{\textsc{[#1]}}}

\usepackage{listings}
\definecolor{TSLKeyword}{HTML}{A626A4}
\definecolor{TSLAttr}{HTML}{005CC5}
\definecolor{TSLComment}{HTML}{6A737D}
\definecolor{TSLString}{HTML}{032F62}
\lstdefinelanguage{TSL}{
  morekeywords={rule,var,where,PROPAGATE,FORWARD,SINK,INJECT,EXTRACT,SET,CLEAR,PRESERVE,COLLAPSE,source,sink,propagate,clear,let,set,inject,extract},
  morekeywords=[2]{@self,@ret,@args,@argc,@is_new,@cwe,value,is_proto,taint,string},
  sensitive=true,
  morecomment=[l]{\#},
  morestring=[b]",
}
\newlist{rqlist}{description}{1}
\setlist[rqlist]{
  style=unboxed,
  font=\normalfont,
  labelsep=0.5em,
  labelwidth=20pt,
  leftmargin=!,
  align=left,
  itemindent=0pt,
  itemsep=0.2em,
  topsep=0.2em
}

\def\BibTeX{{\rm B\kern-.05em{\sc i\kern-.025em b}\kern-.08em
    T\kern-.1667em\lower.7ex\hbox{E}\kern-.125emX}}

\makeatletter
\providecommand{\linebreakand}{%
  \end{@IEEEauthorhalign}%
  \hfill\mbox{}\\%
  \mbox{}\hfill\begin{@IEEEauthorhalign}%
}
\makeatother

\begin{document}

\title{\dsl : Declarative Dynamic Taint Analysis via Shadow Virtual Machine}

\author{
  \IEEEauthorblockN{Zhuohao Zhang}
  \IEEEauthorblockA{\textit{Johns Hopkins University} \\
  Baltimore, MD, USA \\
  zzhan381@jhu.edu}
  \and
  \IEEEauthorblockN{Junkun Liu}
  \IEEEauthorblockA{\textit{Johns Hopkins University} \\
  Baltimore, MD, USA \\
  jliu384@jh.edu}
  \and
  \IEEEauthorblockN{Rui Yang}
  \IEEEauthorblockA{\textit{Johns Hopkins University} \\
  Baltimore, MD, USA \\
  ryang54@jh.edu}
  \linebreakand
  \IEEEauthorblockN{Yinzhi Cao}
  \IEEEauthorblockA{\textit{Johns Hopkins University} \\
  Baltimore, MD, USA \\
  yinzhi.cao@jhu.edu}
  \and
  \IEEEauthorblockN{Ziyang Li}
  \IEEEauthorblockA{\textit{Johns Hopkins University} \\
  Baltimore, MD, USA \\
  ziyang@cs.jhu.edu}
}

\maketitle

\begin{abstract}
Dynamic taint analysis (DTA) for interpreted languages like JavaScript and Python requires three capabilities:
observing host-runtime operations,
maintaining parallel taint states,
and defining how taint propagates across host operations.
Existing systems couple these capabilities within a particular instrumentation mechanism---source-rewriting or engine-native---either incurring high runtime overhead or demanding engine-specific embeddings and representations.
There is yet to be a runtime-independent abstraction of a general DTA that separates taint semantics and state transitions from how a host runtime observes and executes them.

We set out to develop a DTA engine that is extensible, performant, and accurate.
To achieve this goal, we introduce a \emph{Shadow Virtual Machine} executing alongside diverse host runtimes that tracks register-level taint, heap-level taint, provenance, and cross-invocation context.
To drive this machine, we design \underline{\dsl}, a declarative taint specification language with formal operational semantics.
\dsl is designed to be language model friendly, and is equipped with validators that enable trustworthy automated synthesis of rules.
Supporting a new vulnerability class requires only adding declarative rules, with no engine modification.
\dsl is also the first to express higher-order function taint transfer declaratively, bridging native-callback boundary with formal semantics.
Further, \dsl rules are compiled ahead of time to a binary representation and dispatch in constant runtime.

We implement our vision into a tool named \tool, which contains a shared core engine and instantiations of the shadow VM on three runtimes along three orthogonal axes: V8 in both Node.js and Chromium (embedding), SpiderMonkey (engine), and CPython (language).
Accuracy wise, on SecBench.js (493 in-scope CVEs across four CWE categories), our V8 instantiation achieves 95.5\% recall with zero false positives on patched-version testing.
Regarding performance, the runtime overhead of \tool is only 1.85$\times$ over vanilla Node.js on NodeMedic's benchmarks, and is 22.7$\times$ lower than NodeMedic-FINE on identical workloads,
all the while producing 33.2\% higher recall than NodeMedic-FINE in its supported categories.
\end{abstract}


\section{Introduction}
\label{sec:intro}

Dynamic taint analysis (DTA) has become a central technique for detecting security-relevant data flows in modern softwares.
In JavaScript systems, DTA underpins DOM-XSS and injection detection~\cite{domsday2018},
server-side Node.js vulnerability analysis and exploit generation~\cite{staicu2018understanding,nodemedicfine2025,explodejs2025},
and prototype pollution and gadget discovery~\cite{probetheproto2022,kang2025follow}.
In the age of AI-for-security, DTA serves as the key infrastructure for LLM-guided vulnerability detection and proof-of-concept (PoC) synthesis~\cite{zhu2025locus,wang2025contemporary,simsek2025pocgen}.
Across these settings, the common task is to follow selected runtime values through framework code, native APIs, and runtime-managed objects until they reach security-sensitive operations.

Despite the demand, DTA systems for JavaScript and similar interpreted languages remain architecturally fragmented.
Source-rewriting systems~\cite{jalangi2013,karim2018platform,nodemedic2024} instrument JavaScript before execution, avoiding engine modification but introducing a source-transformation boundary: analyzed modules must be available, successfully transformed, and often transpiled through Babel to handle modern syntax.
This approach forgoes native and JIT-compiled execution, incurring order-of-magnitude slowdowns.
Engine-native systems~\cite{Foxhound,panoptichrome2024} modify runtime representations or execution handlers.
They gain runtime visibility, but their shadow state, propagation logic, and instrumentation are implemented together inside one engine.
Neither approach provides a runtime-independent shadow execution model and declarative taint semantics that can be fast and reusable across execution tiers, embeddings, engines, and languages.

The underlying problem is the absence of a runtime-independent execution model for DTA.
Different runtimes expose operations through different bytecodes, stacks, layouts, and compilation strategies, a taint analysis needs the same logical information from each operation: its identity, arguments, result, nesting context, and affected runtime values.
Likewise, the taint transition itself does not inherently depend on whether the host operation was observed in an interpreter or a JIT.
Separating these concerns would allow the observation mechanism to vary while the analysis remain fixed.

We introduce a \emph{Shadow Virtual Machine}, a runtime-independent abstract machine for DTA.
Runtime adapters translate concrete execution into uniform operation-entry and operation-exit events,
while the Shadow VM maintains the analysis state: shadow stack, shadow heap, provenance graph, and persistent context.
This separates \emph{where} an operation is observed from \emph{how} taint state is updated.
We make this separation explicit through a host interface covering operation events, operand addressing, operation identity, object lifecycle, runtime queries, and asynchronous continuation.
As the same operation event model covers interpreted and JIT-compiled execution alike, taint is preserved through JIT-optimized code rather than forcing a deoptimizing fallback to the interpreter, making the engine performant.

While the runtime execution is instrumented by the Shadow VM, the precise taint behaviors of diverse runtimes and libraries are specified in \dsl, a compiled declarative domain-specific language with formal semantics over the Shadow VM.
\dsl's basic syntax allows to describe sources, sinks, propagation, sanitization, and guarded runtime predicates.
Its novel \texttt{inject} and \texttt{extract} actions model native higher-order functions such as \texttt{Array.map} without engine-specific callback policies.
We implement this design in \tool, a multi-lingual DTA framework for JavaScript and Python.
We evaluate \tool on SecBench.js~\cite{secbenchjs}, NodeMedic's benchmark~\cite{nodemedic2024}, and 19 recent vulnerabilities, measuring recall, false positives, end-to-end runtime overhead, and portability. % Performed on JS,
On JS, \tool is
    faster than state-of-the-art DTA engines by at least 22.7$\times$,
    incurring only 1.85$\times$ overhead from the host engine.
Accuracy wise, \tool achieves $95.5\%$ recall on three curated benchmarks,
    reporting zero false positives on patched-version cases~(§\ref{sec:evaluation}).

In summary, this work makes the following contributions:
\begin{itemize}[leftmargin=*, itemsep=1pt, topsep=2pt]
    \item The Shadow VM abstraction and host interface that separate
    runtime observation from taint state and transitions~(\S\ref{sec:shadowvm}).

    \item \dsl, a compiled declarative taint specification language with formal operational
    semantics~(\S\ref{sec:tsl}).

    \item \tool, a Node.js DTA framework which is ported across Chromium, Spidermonkey, and CPython~(\S\ref{sec:implementation}).

    \item An extensive and systematic evaluation of \tool on runtime performance and detection accuracy~(\S\ref{sec:evaluation}).

\end{itemize}

Our tool \tool and the specification language \dsl, including their instantiations on the three interpreters, are made publicly available at \url{https://github.com/MM0n5Ter/Mystra}.

\section{Motivating Example}

\begin{figure}[t]
\begin{subfigure}[t]{\columnwidth}
\includegraphics[width=\linewidth]{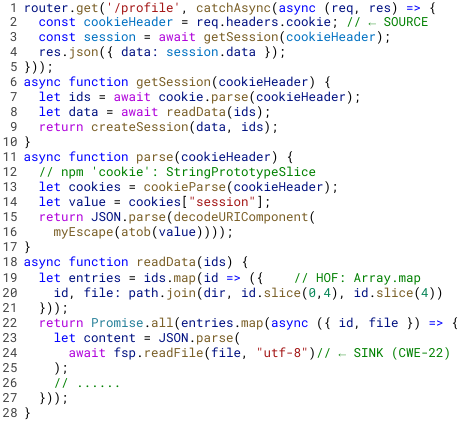}
\caption{Simplified code pattern of CVE-2025-61686.}
\label{fig:motivating-flow}
\end{subfigure}

\begin{subfigure}[t]{\columnwidth}
\begin{lstlisting}
rule buffer.atob(str):
    propagate str -> @ret
rule JsonParse(str):
    propagate str -> @ret
rule ArrayMap(self, callback):
    inject @self[*] -> callback.param[0]
    extract callback.ret -> @ret[*]
rule openFileHandle(_, path):
    sink path @cwe(22)
\end{lstlisting}
\vspace{-10px}
\caption{\dsl rules covering the flow.}
\label{fig:motivating-rules}
\end{subfigure}
\caption{Motivating example: data flow and taint specification.}
\vspace{-10px}
\label{fig:motivating}
\end{figure}

\subsection{Vulnerability Context}
\label{sec:2.1}

We introduce the capability of \tool and \dsl through the detection of a vulnerability, CVE-2025-61686~\cite{reactroutercve}, found in React-router, which is the most widely used routing framework in the Node.js ecosystem~\cite{reactrouterdownload}. 
The vulnerability affects its server-side session storage API: \texttt{createFileSessionStorage} uses an unsigned cookie value as a filesystem path.
The attacker could set session cookie to path traversal payload and use it to read arbitrary files without authentication.

Figure~\ref{fig:motivating} shows the simplified data flow from HTTP input to file-system read.
From the application level, a single \texttt{await getSession(cookieHeader)} call hides seven native operations across four trust domains.
The cookie decoder chains four operations in a single expression: a third-party library call (\texttt{cookieParse}), a Node.js native call (\texttt{atob}), and two V8 runtime intrinsics calls (\texttt{decodeURIComponent}, \texttt{JSON.parse}).
The file storage uses \texttt{Array.map} to process session identifiers through \texttt{path.join} before reaching the sink, \texttt{readFile} (line 24), via \texttt{Promise.all}.
The entire chain executes asynchronously: taint must survive across multiple \texttt{await} boundaries to reach the final \texttt{readFile} call.

\subsection{Challenges and Solutions}
\label{sec:2.2}

\paragraph{Compound operations demanding expressive taint semantics}
Native and runtime-managed operations do not always induce simple argument-to-return dependencies.
Some invoke user callbacks or reconstruct containers, transferring taint in ways a fixed propagation table cannot describe.
Existing DTA engines either encode such behavior as imperative, framework-specific models~\cite{karim2018platform,nodemedic2024},
or hardcode taint propagation inside the native implementation~\cite{panoptichrome2024}.
\tool instead describes operation-specific taint behavior declaratively over a uniform operation model expressed in a DSL.
In \dsl, we employ rules like \texttt{inject} and \texttt{extract} to express how taint transfer across callbacks in native higher-order functions such as \texttt{Array.map} (Fig.~\ref{fig:motivating-rules}, lines~5-7).

\paragraph{Anchoring taint semantics to a uniform execution model}
Expressive rules are only useful if they attach to a faithful, uniform execution model.
Source-rewriting systems define taint over a transformed program, so the operations visible to their rules are not the operations that run: 
Babel's desugaring procedure rewrites callback structure, async scheduling, property accesses, and exception paths before analysis begins~\cite{jalangi2013,nodemedic2024}.
Engine-native systems observe real execution but bind their taint logic to implementation-specific program points, so the semantics are inseparable from the engine's internals~\cite{Foxhound,panoptichrome2024}.
\tool separates observation from semantics: a runtime adapter instruments execution while the taint semantics live in rules defined over an event model with formal operational semantics.
For instance, \tool handles the asynchronous functions in the motivating example (Fig.~\ref{fig:motivating-flow}, lines~1, 6, 11, and 18) as a fixed continuation transition, which NodeMedic reports as a known limitation~\cite{nodemedic2024}.
The rules are anchored to operations, not to rewritten syntax or engine-specific hook code, so the same semantics apply unchanged across interpreted, native, and JIT execution.

Together, \tool turns taint behavior from instrumentation code into a portable semantics: visible as operation events, specified as declarative rules, and executed by a Shadow VM.

\section{Shadow Virtual Machine}
\label{sec:shadowvm}

\begin{figure*}[t]
    \centering
    \includegraphics[width=\textwidth]{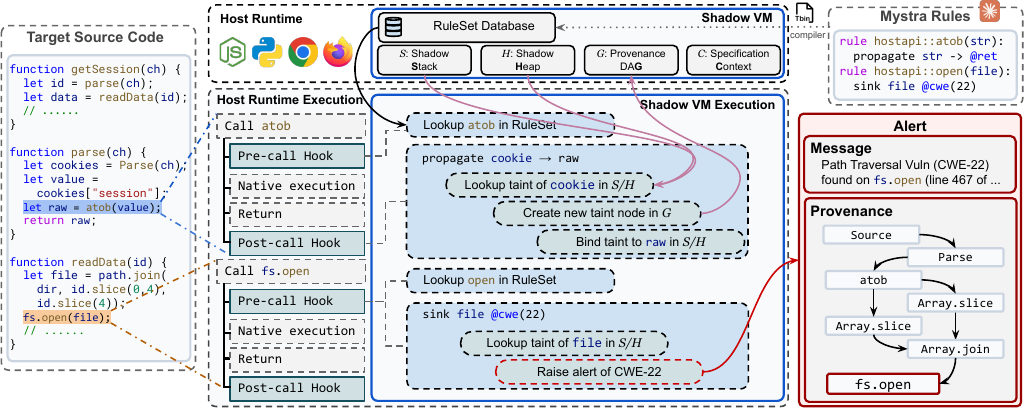}
    \caption{Overview of the proposed architecture.}
    \label{fig:architecture-wide}
    \vspace{-10px}
\end{figure*}

We present the Shadow VM, a parallel virtual machine that tracks taint alongside a host runtime (Figure~\ref{fig:architecture-wide}).
A host adapter projects concrete execution onto operation boundaries, and the Shadow VM updates the parallel state $\sigma=\langle S,H,G,C\rangle$ through generic hook and bridge transitions with rules.
We formally define these states in Figure~\ref{fig:formal-defs}.

\begin{figure}[t]
    \centering
    \footnotesize
    \vspace{-4pt}
    \small
    \[
    \renewcommand{\arraystretch}{0.75}
    \begin{array}{rrcl}
        \text{(Shadow Stack)} & S & \in & \mathit{Frame}^{*} \\
        \text{(Shadow Heap)}  & H & : & \mathit{Addr} \times \mathit{Key} \to \mathit{TaintId} \\
        \text{(Prov. DAG)}    & G & : & \mathit{List}\langle\mathit{FlowNode}\rangle \\
        \text{(Spec. Context)} & C & : & \mathit{VarName} \times \mathit{Val}^{*} \rightharpoonup \mathit{Val} \\
        \text{(Values)} & \mathit{Val} & \in & \mathit{TaintId} \cup \mathit{Ref} \cup \mathbb{Z} \cup \mathit{Str} \cup \mathit{Bool} \\
    \end{array}
    \]
    \caption{Shadow VM state domains.}
    \label{fig:formal-defs}
    \vspace{-16px}
\end{figure}

\subsection{Abstraction}
\label{sec:3.1}

Every host operation is modeled uniformly as $\lambda(\mathit{args})\to\mathit{result}$.
An entry event exposes the operation and arguments, and the matching exit event exposes its result.
Internal execution may contain nested operation events.
The Shadow VM therefore observes the same abstraction regardless of whether the host executes source code, bytecode, or machine code.

Whether an operation requires a taint rule follows from this observability.
A transparent operation exposes its data flow through nested operations and needs no summary.
An opaque operation hides some data flow:
universal operations such as arithmetic and property access use fixed transitions, while runtime- and library-specific operations use \dsl rules.
Both follow the same entry--exit lifecycle.
For example, a native higher-order function is opaque internally, but its user defined callbacks may remain visible as nested operations.

\subsection{Shadow States}
\label{sec:3.2}
\subsubsection{Shadow Stack}

The shadow stack $S$ stores active operation frames in event-nesting order: entry pushes a frame and the matching exit pops it.
A frame maps operands to taint identifiers and stores higher-order bridge metadata.
For transparent functions, operands correspond to runtime registers or stack slots.
Implementations may elide frames for atomic operations while preserving these transitions.

Normal returns pop the current shadow frame, but exceptions and asynchronous suspension may transfer control non-locally.
Before processing subsequent operation events, the adapter synchronizes $S$ with the host’s logical continuation.
Exception synchronization discards unwound frames and installs the thrown-value taint in the handler activation;
asynchronous synchronization restores the saved operand-taint environment of a suspended activation.

\subsubsection{Shadow Heap}

The shadow heap $H$ tracks object properties that persist across operation frames.
It maps object-address and property-key pairs to taint identifiers.
Per-key tracking avoids tainting an entire object from one field; wildcard keys represent unobservable element placement.
Entries follow host-object lifetime through the GC adapter.
Taint resolution consults $S$ first and uses $H$ as fallback.

\subsubsection{Provenance DAG}
While $S$ and $H$ record where taint resides, the provenance DAG $G$ records its origin.
When an operation derives taint, the Shadow VM appends a flow node whose parents reference its operand taints.
Traversing these edges from an alert reconstructs its source-to-sink provenance.

\subsubsection{Specification Context}
State in $S$ and $H$ is tied to active operations and live objects, so taint is lost across external boundaries such as file I/O.
The specification context $C$ provides named stores that persist across such boundaries.
Each store maps a typed key tuple to a value:
$C(x,k_1,\ldots,k_n)\rightharpoonup\mathit{Val}$.
We describe its language constructs in \S\ref{sec:tsl-context}.

\subsection{Host Interface}
\label{sec:3.3}

The interaction between the Shadow VM and the host runtime is mediated by an adapter that implements a single host interface of six capabilities:

\begin{enumerate}[leftmargin=*, itemsep=1pt, topsep=2pt]
  \item \textbf{Operation events.} Well-bracketed entry and exit boundaries, including nested and reentrant operations.
  \item \textbf{Operand addressing.} Mapping arguments, returns, registers, and stack slots to shadow locations.
  \item \textbf{Operation identity.} Stable identities for rule dispatch and higher-order bridge matching.
  \item \textbf{Object lifecycle.} Stable heap identities plus relocation and finalization notifications.
  \item \textbf{Runtime queries.} Values and metadata required by locators, references, and guards.
  \item \textbf{Continuation synchronization.} Identifying non-local control transfers and restoring the target activation’s live shadow operands.
\end{enumerate}

The shared core contains the taint engine, rule interpreter, DAG, and specification context.
Adapters, operation bindings, and runtime-specific rules form the porting surface.
We describe three instantiations in \S\ref{sec:rq4}.

\section{Language Design}
\label{sec:tsl}
We provide an overview of \dsl, the declarative taint specification language we previously illustrated in Figure~\ref{fig:motivating-rules}.
Here, we illustrate each key construct using examples of taint rules for common JavaScript operations.

\subsection{Rules and Actions}
A taint analysis begins with two questions: where does untrusted data enter, and where is it dangerous. \dsl introduces \texttt{source} and \texttt{sink} to define these boundaries as shown in the following two rules. Each rule begins with the \texttt{rule} keyword and a namespaced signature using a \texttt{::} to separate a namespace from an actual function signature. Named parameters bind to arguments in the function signature.
The arrow \texttt{->} separates sources from destinations and keywords prefixed with \texttt{@} are built-in constructs of the language.

\vspace{-6px}
\begin{lstlisting}
rule V8_hostapi::CreateHTTPServer.request:
    source -> @ret
rule V8_native::openFileHandle(_, path):
    sink path @cwe(22)
\end{lstlisting}
\vspace{-6px}

\noindent The first rule marks the return value of the function \texttt{request} as a taint source (line 2): the action \texttt{source} creates a fresh taint and writes it to the return value \texttt{@ret}. 
The second rule checks whether argument \texttt{path} carries taint when the function \texttt{openFileHandle} is called. If so, an alert of path traversal is raised as annotated by \texttt{@cwe(22)} (line 4). Together, these two rules define a minimal end-to-end detection.

Between source and sink, taint must propagate through intermediate operations. Action \texttt{propagate} models this flow, and \texttt{clear} models sanitization:

\vspace{-6px}
\begin{lstlisting}
rule V8_native::StringPrototypeConcat(...parts):
    propagate @self, parts -> @ret
rule V8_native::StringPrototypeStartsWith:
    clear
\end{lstlisting}
\vspace{-6px}

\noindent The first rule reads: the return value \texttt{@ret} is tainted if either the receiver \texttt{@self} or any of the variadic arguments \texttt{parts} carries taint. The second rule uses \texttt{clear}: \texttt{startsWith} returns a boolean not derived from the input content, so taint is removed.

A rule can contain multiple actions and conditional guards. The following rule for \texttt{Object.defineProperty} combines propagation with a guarded sink:

\vspace{-6px}
\begin{lstlisting}
rule V8_native::ObjectDefineProperty(tar, key, desc):
    propagate desc.value -> tar[key]
    sink desc @cwe(1321) where tar.is_proto
\end{lstlisting}
\vspace{-6px}

\noindent The \texttt{propagate} reads taint from \texttt{desc.value}
and writes it to \texttt{tar[key]}, where the bracketed \texttt{key} is resolved at runtime to the actual property name being defined.
The \texttt{sink} raises a CWE-1321 alert if \texttt{tar} is a prototype object. The \texttt{where} clause conditions an action on a runtime predicate, preventing false alerts on non-dangerous cases.

\newcommand{\kw}[1]{\texttt{#1}} 

\begin{figure*}[t]
\centering
\small
% --- Left: Structure + Actions ---
\begin{minipage}[t]{0.48\textwidth}
    \centering
    \textbf{Structure and Actions} \\[0.2em]
    $
    \renewcommand{\arraystretch}{0.75}
    \begin{array}{rrcl}
        \text{(Program)}  & P & ::= & \overline{d} \;;\; \overline{r} \\[1.5pt]
        \text{(Var Decl)} & d & ::= & \kw{let}\; x\; [\overline{\tau}] ~\texttt{->}~ \tau \\[1.5pt]
        \text{(Rule)}     & r & ::= & \kw{rule}\; \mathit{ns}{::}\mathit{sig}(\overline{p}) {:}\; \overline{s} \\
        \text{(Sub-rule)} & s & ::= & a \;\;[\; g \;] \\
        \text{(Action)}   & a & ::= & \kw{propagate}\; \overline{\ell} ~\texttt{->}~ \overline{\ell} \\
                          &   & \mid & \kw{source} ~\texttt{->}~ \overline{\ell} \\
                          &   & \mid & \kw{sink}\; \overline{\ell}\;\; \texttt{@cwe}(n) \\
                          &   & \mid & \kw{inject}\; \overline{\ell} ~\texttt{->}~ \mathit{f}\texttt{.param}[i] \\
                          &   & \mid & \kw{extract}\; \mathit{f}\texttt{.ret} ~\texttt{->}~ \overline{\ell} \\
                          &   & \mid & \kw{set}\; x[\overline{v}] \;\texttt{=}\; v \\
                          &   & \mid & \kw{clear} \\[2pt]
        \text{(Value)}    & v & ::= & \ell\texttt{.taint} \mid \ell\texttt{.ref} \mid \ell\texttt{.value} \\
                          &   & \mid & c \mid x[\overline{v}] \mid v \oplus v
    \end{array}
    $
\end{minipage}
\hfill
% --- Right: Locators + Predicates ---
\begin{minipage}[t]{0.48\textwidth}
    \centering
    \textbf{Locators and Predicates} \\[0.2em]
    $
    \renewcommand{\arraystretch}{0.75}
    \begin{array}{rrcl}
        \text{(Locator)}   & \ell & ::= & b \;[\;\pi\;] \\[1.5pt]
        \text{(Base)}      & b    & ::= & \texttt{@self} \mid \texttt{@ret} \mid \texttt{@args}[i] \mid p \\[1.5pt]
        \text{(Path)}      & \pi  & ::= & \epsilon \mid \texttt{[*]}  \mid \texttt{.}k \mid \texttt{[}k\texttt{]} \\[1.5pt]
        % \text{(Callback) \ziyang{Delete this}}  & \mathit{cb} & ::= & p\texttt{.param}[i] \mid p\texttt{.ret} \\[3pt]
        %
        \text{(Guard)}     & g    & ::= & \texttt{where}\; \phi \\[1.5pt]
        \text{(Predicate)} & \phi & ::= & e \mid e \;\mathit{op}\; c \mid \phi \;\texttt{and}\; \phi \\
                           &      & \mid & \phi \;\texttt{or}\; \phi \mid \texttt{not}\; \phi \\[1.5pt]
        \text{(Guard Expr)}& e    & ::= & \ell\texttt{.type} \mid \ell\texttt{.is\_fn} \mid \ell\texttt{.is\_proto} \\
                           &      & \mid & \texttt{@argc} \mid \texttt{@is\_ctor} \\[6pt]
        \text{(Namespace)} & \mathit{ns} & \in & \mathit{Namespace} \\[2pt]
        \text{(Type)}      & \tau & \in  & \{ \kw{taint}, \kw{ref}, \kw{str}, \kw{bool}, \kw{num} \} \\[2pt]
        \text{(Constant)}  & c    & \in  & \mathit{Literal} \\[2pt]
        \text{(Identifier)}& p, x, f, k & \in & \mathit{Name}
    \end{array}
    $
\end{minipage}
% \vspace{0.2em}
\caption{Core Abstract syntax of \dsl.
  $\overline{\cdot}$ denotes a sequence.
  $[\;\cdot\;]$ denotes optional.}
\label{fig:tsl-syntax}
\vspace{-10px}
\end{figure*}

\subsection{Higher-order Function}
Higher-order functions are pervasive in modern interpreted languages:
\texttt{map}, \texttt{filter}, and \texttt{reduce} appear in JavaScript, Python, and Ruby alike.
In an operation such as \texttt{Array.map}, the native loop and result assembly are opaque, while each user-function invocation remains visible as a nested operation.
\dsl provides \texttt{inject} and \texttt{extract} to bridge taint across these nested boundaries:

\vspace{-6px}
\begin{lstlisting}
rule V8_native::ArrayPrototypeMap(callback, arg):
    inject @self[*] -> callback.param[0]
    extract callback.ret -> @ret[*]
\end{lstlisting}
\vspace{-6px}

\noindent Here, \texttt{callback} is the rule parameter bound to the user function. \texttt{inject} registers a bridge from the container's element taint \texttt{@self[*]} to its first parameter.
Each matching entry realizes this transfer, and matching exits accumulate return taints.
When \texttt{Array.map} exits, \texttt{extract} writes the merged taint to \texttt{@ret[*]}.
The user function itself is transparent and needs no declarative summary. 
We formalize this lifecycle in \S\ref{sec:tsl-semantics}.

\subsection{Specification Context Extension}
\label{sec:tsl-context}
Taint in the shadow stack $S$ and heap $H$ is tied to live objects and active operation frames. When a tainted value crosses an invocation boundary, the connection is lost: \texttt{writeFile} and a subsequent \texttt{readFile} execute in different frames with no surviving taint.
\dsl provides specification context variables in $C$ to bridge these gaps:

\vspace{-6px}
\begin{lstlisting}
let filelist[string] -> taint;
rule writeFile(path, data):
    set filelist[path.value] = data.taint
rule readFile(path):
    propagate filelist[path.value] -> @ret
\end{lstlisting}
\vspace{-6px}

\noindent The \texttt{let} action declares a global storage space \texttt{filelist} keyed by string and mapping to datatype taint.
When \texttt{writeFile} is called, \texttt{set} records the taint of \texttt{data} under the \texttt{path}.
When \texttt{readFile} is later called with the same path, propagate recovers the stored taint into \texttt{@ret}.  
Variable key and value types are checked at compile time, preventing mismatched lookups.

\subsection{Operational Semantics}
\label{sec:tsl-semantics}

% ==================================================================
% Event-driven operation semantics with callback bindings in shadow frames.
%
% Every operation follows the same lifecycle:
%   enter -> internal execution -> exit.
% Internal execution may contain nested operations.  Hence callbacks are not
% a separate semantic category: they are ordinary nested operations whose
% frames consult the immediately preceding frame for matching INJECT records.
% ==================================================================
\begin{figure*}[t]
% \centering
\footnotesize
\setlength{\abovedisplayskip}{3pt}
\setlength{\belowdisplayskip}{3pt}
\setlength{\abovedisplayshortskip}{2pt}
\setlength{\belowdisplayshortskip}{2pt}
\setlength{\jot}{2pt}

% --- Notation ---
\noindent\textbf{Notation.}\quad
$\sigma=\langle S,H,G,C\rangle$;
$\mathbb{R}$ is the host-runtime state;
$\mathbb{E}$ is the host-event stream;
$\Gamma$ is the compiled rule library.
$\mathcal{R}$ and $\mathcal{W}$ resolve and write taint;
$\mathit{newnode}(T,G)$ creates a provenance node with parents $T$.
$\mathit{register}$, $\mathit{bridgein}$, $\mathit{bridgeout}$, and
$\mathit{accum}$ implement HOF bridging.

% --- Every operation reduces through entry, internal execution, and exit. ---
\fbox{
\smallskip\noindent\textbf{Augmented Reduction.}\quad
$\mathbb{R},e::\mathbb{E}
  \crarrow
  \mathbb{R}',\mathbb{E}$: original host language reduction; 
$\mathbb{R},\tau::\mathbb{E},\sigma
  \cRarrow
  \mathbb{R}',\mathbb{E},\sigma$: the augmented reduction with Shadow VM.
}
\[
(\text{Event})~~e ::= \mathsf{enter}(op,\bar a)
  \mid \tau
  \mid \mathsf{exit}(op,v)
\qquad
  \mathit{lookup}_{\Gamma}(op)=
  \begin{cases}
    \bar r,
      & \text{if}\langle op,\bar r\rangle\in\Gamma,\\
    \epsilon,
      & \text{otherwise.}
  \end{cases}
\]
\begin{minipage}[b]{0.32\textwidth}
\[
\frac{
  \begin{array}{c}
  \mathbb{R},op,\sigma\vdash
    \mathit{prehook}(\bar a)
    \cDownarrow\sigma'
  \\[2pt]
  \mathbb{R},\mathsf{enter}(op,\bar a)::\mathbb{E}
  \crarrow
  \mathbb{R}',\mathbb{E}
  \end{array}
}{
  \mathbb{R},
  \mathsf{enter}(op,\bar a)::\mathbb{E},
  \sigma
  \cRarrow
  \mathbb{R}',\mathbb{E},\sigma'
}
\;\;\rulename{E-Enter}
\]
\end{minipage}
\hfill
\begin{minipage}[b]{0.32\textwidth}
\[
\frac{
  \mathbb{R},\tau::\mathbb{E}
  \crarrow
  \mathbb{R}',\mathbb{E}
}{
  \mathbb{R},\tau::\mathbb{E},\sigma
  \cRarrow
  \mathbb{R}',\mathbb{E},\sigma
}
\;\;\rulename{E-Internal}
\]
\end{minipage}
\hfill
\begin{minipage}[b]{0.32\textwidth}
\[
\frac{
  \begin{array}{c}
  \mathbb{R},op,\sigma\vdash
    \mathit{posthook}(v)
    \cDownarrow\sigma'
  \\[2pt]
  \mathbb{R},
  \mathsf{exit}(op,v)::\mathbb{E}
  \crarrow
  \mathbb{R}',\mathbb{E}
  \end{array}
}{
  \mathbb{R},
  \mathsf{exit}(op,v)::\mathbb{E},
  \sigma
  \cRarrow
  \mathbb{R}',\mathbb{E},\sigma'
}
\;\;\rulename{E-Exit}
\]
\end{minipage}

\smallskip

% --- Hook evaluation ---
\fbox{
\noindent\textbf{Hook Evaluation.}\quad
$\mathbb{R},op,\sigma\vdash h \cDownarrow\sigma'$, $h$ is a hook.
}
\[
  \begin{array}{l}
  \mathit{bridgein}(op,\sigma)=
    \sigma[S_{\mathit{top}}.\mathit{param}[i]\mapsto T]
    \quad
    \text{for each } \langle op,i,T\rangle
    \in S_{\mathit{parent}}.\mathit{bridges}
  \\[3pt]
  \mathit{bridgeout}(op,\sigma)=
    \sigma[S_{\mathit{parent}}.\mathit{acc}[op]\mapsto
    S_{\mathit{parent}}.\mathit{acc}[op]\cup
    \mathcal{R}(\texttt{@ret},\sigma)]
    \quad
    \text{for each } \langle op,i,T\rangle
    \in S_{\mathit{parent}}.\mathit{bridges}
  \end{array}
\]
\[
  \frac{
    \bar r=\mathit{lookup}_{\Gamma}(op)
    \quad
    \sigma'=\sigma[
      S\mapsto\mathit{push}(
        \sigma.S,\mathit{newframe}(\sigma,\bar a))]
    \quad
    \sigma''=\mathit{bridgein}(op,\sigma')
    \quad
    \mathbb{R},op,\sigma''\vdash
      \mathit{pre}(\bar r)
      \cDownarrow\sigma'''
  }{
    \mathbb{R},op,\sigma\vdash
      \mathit{prehook}(\bar a)
      \cDownarrow\sigma'''
  }
  \;\;\rulename{Pre-Hook}
\]
\[
  \frac{
    \bar r=\mathit{lookup}_{\Gamma}(op)
    \quad
    \mathbb{R},op,\sigma\vdash
      \mathit{post}(\bar r,v)
      \cDownarrow\sigma'
    \quad
    \sigma''=\mathit{bridgeout}(op,\sigma')
    \quad
    \sigma'''=\sigma''[
      S\mapsto\mathit{pop}(\sigma''.S)]
  }{
    \mathbb{R},op,\sigma\vdash
      \mathit{posthook}(v)
      \cDownarrow\sigma'''
  }
  \;\;\rulename{Post-Hook}
\]

\smallskip

% --- Rule Evaluation ---
\fbox{
\noindent\textbf{Rule Evaluation.}\quad
$\mathbb{R},op,\sigma\vdash\bar r\cDownarrow\sigma'$, $\bar r$ is a set of rules.
}
\[
\frac{}{
  \mathbb{R},op,\sigma\vdash\epsilon\cDownarrow\sigma
}
\;\;\rulename{Empty}
\qquad
\frac{
  \mathbb{R},op,\sigma\vdash r\cDownarrow\sigma'
  \quad
  \mathbb{R},op,\sigma'\vdash\bar r\cDownarrow\sigma''
}{
  \mathbb{R},op,\sigma\vdash(r::\bar r)\cDownarrow\sigma''
}
\;\;\rulename{Seq}
\]

\[
  \frac{
  eval(\phi, \sigma, \mathbb{R}) = \top \quad 
  \mathbb{R},op,\sigma\vdash a\cDownarrow\sigma'
  }{
  \mathbb{R},op,\sigma\vdash(a\;\texttt{where}\;\phi)\cDownarrow\sigma'
  }
\;\;\rulename{Guard-T}
\qquad
  \frac{
  eval(\phi, \sigma, \mathbb{R}) = \bot
  }{
  \mathbb{R},op,\sigma\vdash(a\;\texttt{where}\;\phi)\cDownarrow\sigma
  }
\;\;\rulename{Guard-F}
\]

% \[
%   \frac{
%   \mathbf{if}\ \phi\ \mathbf{then}\
%   (\mathbb{R},op,\sigma\vdash a\cDownarrow\sigma')
%   \ \mathbf{else}\
%   (\sigma'=\sigma)
%   }{
%   \mathbb{R},op,\sigma\vdash(a\;\texttt{where}\;\phi)\cDownarrow\sigma'
%   }
% \;\;\rulename{Guard}
% \]
% \end{minipage}

\smallskip

% --- Per-Action Rules ---
\fbox{
\noindent\textbf{Action Evaluation.}\quad
$\mathbb{R},op,\sigma\vdash r\cDownarrow\sigma'$, $r$ is a single rule.
}
% --- PROPAGATE ---
\[
% \frac{
%   T=
%   \bigcup_{\ell\in\overline{\ell}_{\mathit{src}}}
%     \mathcal{R}(\ell,\sigma)
%   \quad
%   T = \emptyset
% }{
%   \mathbb{R},op,\sigma\vdash
%   r
%   \cDownarrow
%   \sigma
% }
% \;\;\rulename{E-Empty}
% \quad
\frac{
  T=
  \bigcup_{\ell\in\overline{\ell}_{\mathit{src}}}
    \mathcal{R}(\ell,\sigma)
  \quad
  T\neq\emptyset
  \quad
  n=\mathit{newnode}(T,G)
  \quad
  \langle S',H'\rangle=
  \mathcal{W}(
    \overline{\ell}_{\mathit{dst}},
    n,
    \langle S,H\rangle)
}{
  \mathbb{R},op,\langle S,H,G,C\rangle\vdash
  \texttt{propagate}\;\overline{\ell}_{\mathit{src}}
    ~\texttt{->}~ \overline{\ell}_{\mathit{dst}}
  \cDownarrow
  \langle S',H',G\cup\{n\},C\rangle
}
\;\;\rulename{E-Propagate}
\]

% --- SOURCE and CLEAR ---
\begin{minipage}{0.52\textwidth}
\[
\frac{
  n=\mathit{newnode}(\emptyset,G)
  \quad
  \langle S',H'\rangle=
  \mathcal{W}(
    \overline{\ell}_{\mathit{dst}},
    n,
    \langle S,H\rangle)
}{
  \mathbb{R},op,\langle S,H,G,C\rangle\vdash
  \texttt{source} ~\texttt{->}~ \overline{\ell}_{\mathit{dst}}
  \cDownarrow
  \langle S',H',G\cup\{n\},C\rangle
}
\;\;\rulename{E-Source}
\]
\end{minipage}
\hfill
\begin{minipage}{0.42\textwidth}
\[
\frac{
  \langle S',H'\rangle=
  \mathcal{W}(\texttt{@ret},0,\langle S,H\rangle)
}{
  \mathbb{R},op,\langle S,H,G,C\rangle\vdash\texttt{clear}
  \cDownarrow
  \langle S',H',G,C\rangle
}
\;\;\rulename{E-Clear}
\]
\end{minipage}

% --- SINK and SET ---
\begin{minipage}{0.55\textwidth}
\[
\frac{
  T=
  \bigcup_{\ell\in\overline{\ell}_{\mathit{src}}}
    \mathcal{R}(\ell,\sigma)
  \quad
  T\neq\emptyset
  \quad
  \mathit{Alert}(c,T)
}{
  \mathbb{R},op,\sigma\vdash
  \texttt{sink}\;\overline{\ell}_{\mathit{src}}\;\texttt{@cwe}(c)
  \cDownarrow\sigma
}
\;\;\rulename{E-Sink}
\]
\end{minipage}
\hfill
\begin{minipage}{0.4\textwidth}
\[
\frac{
  \overline{k}=
  \mathit{eval}(\overline{v}_{\mathit{key}},\sigma,\mathbb{R})
  \quad
  u=\mathit{eval}(v_{\mathit{rhs}},\sigma,\mathbb{R})
}{
  \mathbb{R},op,\sigma\vdash
  \texttt{set}\;x[\overline{v}_{\mathit{key}}] ~\texttt{=}~ v_{\mathit{rhs}}
  \cDownarrow
  \sigma[C[x,\overline{k}]\mapsto u]
}
\;\;\rulename{E-Set}
\]
\end{minipage}

% --- INJECT: store the callback binding in the current frame. ---
\[
  \frac{
    T=
    \bigcup_{\ell\in\overline{\ell}_{\mathit{src}}}
      \mathcal{R}(\ell,\sigma)
    \quad
    op_f=\mathit{Ref}(f,\mathbb{R})
    \quad
    S'=\mathit{register}
      (S,\langle op_f,i,T\rangle)
  }{
    \mathbb{R},op,\langle S,H,G,C\rangle\vdash
    \texttt{inject}\;\overline{\ell}_{\mathit{src}}
      ~\texttt{->}~ f\texttt{.param}[i]
    \cDownarrow
    \langle S',H,G,C\rangle
  }
  \;\;\rulename{E-Inject}
\]

% --- EXTRACT: merge returns collected in the current frame. ---
\[
  \frac{
    op_f=\mathit{Ref}(f,\mathbb{R})
    \quad
    T=\mathit{accum}(S,op_f)
    \quad
    T\neq\emptyset
    \quad
    n=\mathit{newnode}(T,G)
    \quad
    \langle S',H'\rangle=
    \mathcal{W}(
      \overline{\ell}_{\mathit{dst}},
      n,
      \langle S,H\rangle)
  }{
    \mathbb{R},op,\langle S,H,G,C\rangle\vdash
    \texttt{extract}\;f\texttt{.ret}
      ~\texttt{->}~\overline{\ell}_{\mathit{dst}}
    \cDownarrow
    \langle S',H',G\cup\{n\},C\rangle
  }
  \;\;\rulename{E-Extract}
  \]

\vspace{-2pt}
\caption{Core operational semantics of \dsl calls and actions; actions are treated as no-ops when there is no taint.}
\label{fig:tsl-semantics}
\vspace{-10px}
\end{figure*}

We formalize \dsl over host execution state ($\mathbb{R}$) and a host-generated event stream ($\mathbb{E}$) (Figure~\ref{fig:tsl-semantics}).
Each dynamic operation is divided into an entry event, a finite body event sequence, and an exit event.
The entry event occurs after the operation and its arguments are resolved but before its body executes; the pre-hook prepares and pushes the corresponding shadow frame.
The exit event occurs after the result is produced but before control returns to the enclosing operation; the post-hook processes the result and pops the frame.
A $\tau$ event represents host execution that requires no Shadow VM state transition.
The body sequence may contain nested operation events and $\tau$ events.

\paragraph{Augmented reduction.}
\textsc{E-Enter}, \textsc{E-Internal}, and \textsc{E-Exit} synchronize host execution with the Shadow VM.
At entry, the pre-hook runs immediately before the host transfers control into the operation. 
Internal execution advances only $\mathbb{R}$. 
At exit, the result $v$ is available to the post-hook, which runs immediately before the host returns to the enclosing operation.
Each reduction consumes exactly the head of $\mathbb{E}$; the Shadow VM does not generate or rewrite the remaining event stream.

\paragraph{Hook evaluation.}
\textsc{Pre-Hook} looks up the operation's rule, creates a frame for its arguments, and pushes it onto $S$.
The entry bridge then transfers any registered taint from the immediately enclosing frame into parameters of the new frame, after which the rule's pre-actions execute.
\textsc{Post-Hook} executes post-actions while the current frame and result remain addressable.
The exit bridge then transfers the current operation's return taint into a matching accumulator in the immediately enclosing frame, and the current frame is popped.
Thus every activation follows
$\mathit{push}\rightarrow\mathit{bridgein}\rightarrow\mathit{pre}
\rightarrow\text{execution}\rightarrow\mathit{post}
\rightarrow\mathit{bridgeout}\rightarrow\mathit{pop}$.

\paragraph{Rule evaluation.}
A rule body is a sequence of guarded actions.  \textsc{Empty} leaves the state unchanged, and \textsc{Seq} threads it through actions from left to right.
\textsc{Guard} evaluates an action only when its \texttt{where} predicate holds; otherwise it preserves the incoming state.
The judgment carries $\mathbb{R}$ and the current operation as read-only context, allowing guards and actions to query runtime values and properties.

\begin{algorithm}[t]
\caption{Hierarchical Taint Resolution}
\label{alg:resolve}
\footnotesize
\begin{algorithmic}[1]
\Require Locator $\ell$; host state $\mathbb{R}$; shadow state
  $\sigma=\langle S,H,G,C\rangle$
\Ensure Taint set $T$
\If{$\ell = x[\overline{v}]$ \textbf{and} $x \in \mathit{Globals}$}
  \State $\overline{k} = \mathit{eval}(\bar{v},\sigma,\mathbb{R})$
  \State \Return $\{\, C[x, \overline{k}] \mid C[x,\overline{k}]\neq 0 \,\}$
\EndIf
\State $q \gets \mathit{top}(S)$ \Comment{active operation frame}
\State $b \gets \text{base}(\ell)$
\State $t_S \gets S[q,\, b]$ \Comment{current-frame lookup}
\State $t_H \gets H[\mathit{addr}(b),\, \texttt{*}]$ \Comment{shadow heap fallback}
\State $t \gets \textbf{if}\ t_S \neq 0\ \textbf{then}\ t_S\ \textbf{else}\ t_H$
\If{$\ell = b$}
  \State \Return $\{\,t\mid t\neq 0\,\}$
\ElsIf{$\ell = b\texttt{.}k$ \textbf{or} $\ell = b[k]$}  
  \State \Return $\{\, H[\mathit{addr}(b),\, k]
    \mid H[\mathit{addr}(b),k]\neq 0 \,\}$
\ElsIf{$\ell = b\texttt{[*]}$}
  \State \Return $\{\, H[\mathit{addr}(b),\, *]
    \mid H[\mathit{addr}(b),*]\neq 0 \,\}$
\EndIf
\end{algorithmic}
\end{algorithm}

\paragraph{Action rules.}
Each action transforms only shadow state under the read-only context $\langle\mathbb{R},\mathit{op}\rangle$.
The resolution function $\mathcal{R}(\ell,\sigma)$ (Algorithm~\ref{alg:resolve}) reads from the active
frame at the top of $S$, using the shadow heap as fallback. 
The write function $\mathcal{W}(\ell,t,\langle S,H\rangle)$ updates the active frame or shadow heap according to the destination locator.
Both use $\mathbb{R}$ as an implicit read-only context for concrete references, values, and dynamic keys.

The action rules follow their surface syntax: \textsc{E-Propagate} merges source taints into a fresh provenance node, \textsc{E-Source} creates a parentless node, \textsc{E-Sink} emits an alert without changing state, \textsc{E-Clear} clears \texttt{@ret}, and \textsc{E-Set} updates $C$.
\textsc{E-Inject} registers a HOF bridge, and \textsc{E-Extract} flushes the bridge accumulator to its destination.
Empty taint sets are no-ops, except that inject still registers the function identity so taint created inside the nested function can be extracted.

Thus, \dsl provides a trace-relative guarantee: for any event trace satisfying the portability contract, if the loaded rules conservatively summarize the explicit dependencies of all opaque operations encountered, then all modeled explicit source-to-sink flows are propagated to sinks.

\subsection{Compilation}
\label{sec:tsl-compilation}

\dsl rules are compiled ahead of time into a binary representation (\texttt{.tbin}) loaded once at VM startup.
The compiler type-checks locators, guards, parameter names, and specification-context keys/values, then schedules actions by action type and locator availability into pre- and post-hook instruction lists.
\texttt{sink} and \texttt{inject} execute before the host operation;
\texttt{clear} and \texttt{extract} execute after the result is available;
\texttt{source}, \texttt{propagate}, and \texttt{set} are placed at the earliest phase in which their referenced locators can be resolved.
Within each phase, actions of the same type retain source order.
The emitted binary uses builtin-ID arrays for native functions and host-signature hash maps for host APIs, giving O(1) dispatch with no runtime parsing or string comparison.

\section{Implementation}
\label{sec:implementation}
We instantiate the Shadow VM in V8 through both the Ignition interpreter and Maglev JIT.
The implementation comprises a runtime-independent core---taint engine, \dsl interpreter, rule loader, provenance graph, and specification context---from a V8 adapter satisfying the portability contract (\S\ref{sec:3.3}).
Ignition hooks and Maglev IR nodes realize the same operation events and Shadow VM transitions; the JIT introduces no separate taint semantics.

\noindent\textbf{Shadow-state realization.}
The shadow stack $S$ is a pre-allocated stack of fixed-size frames storing operand taints and higher-order bridge metadata.
Ignition addresses operands by register index; Maglev realizes the same logical locations through compiled shadow operations. 
The shadow heap $H$ is an external hash map keyed by object identity.
Relocation and finalization hooks maintain its entries as objects move or die.
The append-only provenance graph $G$ and typed specification context $C$ reside outside V8's managed heap.

\noindent\textbf{Operation events.}
V8 pre- and post-hooks realize operation entry and exit.
The pre-hook observes the operation and arguments and executes entry-phase actions;
after the result is available, the post-hook executes exit-phase actions.
Nested calls produce nested hook pairs.
Atomic operations such as arithmetic and property access inline both transitions in one bytecode handler and elide the semantic frame.
Calls use a post-call bytecode to match exit transition.

\noindent\textbf{Continuation synchronization.}
Exceptions and asynchronous suspension bypass normally paired operation exits.
The adapter therefore synchronizes $S$ to the resumed host continuation: exception unwinding discards abandoned shadow frames and
transfers the exception taint to the handler, while Promise-associated state restores operand taints after \texttt{await} or callback resumption.
Both preserve $H$, $G$, and $C$.

\noindent\textbf{Tier-independent execution.}
Ignition realizes Shadow VM transitions through bytecode hooks, whereas Maglev lowers the same transitions into compiled IR nodes. 
We introduce 11 Maglev nodes that update the same shadow state and invoke the same rule semantics as the interpreter implementation.
The performance-critical \texttt{DtaCallPreHook} classifies calls using a runtime-function lookup table, a user-function metadata bit, and a builtin bitmap.
Calls requiring no rule work continue in two to three machine instructions without entering C++.
These fast paths keep common untainted execution inline; \S\ref{sec:rq2} evaluates their end-to-end effect.

\noindent\textbf{Other instantiations.}
The V8 instantiation also supports Chromium through DOM-specific rules and host-operation bindings.
SpiderMonkey and CPython reuse the shared core through separate adapters and runtime-specific operation bindings.
We evaluate portability in \S\ref{sec:rq4}.

\section{Evaluation}
\label{sec:evaluation}
\noindent We aim to answer the following research questions:
\begin{rqlist}
  \item[\textbf{RQ1}]
    \textbf{Detection Accuracy:} How accurately does \tool detect vulnerabilities, measured by recall and false positives?
  \item[\textbf{RQ2}]
    \textbf{Performance:} What is the runtime overhead of \tool compared to existing approaches?
  \item[\textbf{RQ3}]
    \textbf{Expressiveness:} Can \dsl express taint behavior of real-world APIs, and can LLMs synthesize valid rules?
  \item[\textbf{RQ4}]
    \textbf{Portability:} Can \tool be ported to a new runtime without modifying its shared core?
\end{rqlist}

\noindent\textbf{Setup.}
All experiments run on a virtual machine with 16-core x86\_64 vCPU, 64GB RAM, Ubuntu 24.04.
We evaluate on SecBench.js~\cite{secbenchjs}, NodeMedic's benchmark~\cite{nodemedic2024} and 19 real-world vulnerabilities disclosed in 2024--2026 spanning the Node.js, CPython, and browser runtimes (Table~\ref{tab:realworld}).
The Shadow VM runs with the full 303-LoC \dsl specification which was manually authored and iteratively refined using API-level tests.
For NodeMedic-FINE, we measure its dynamic taint-analysis stage as the latest maintained NodeMedic rather than its fuzzing and exploit-synthesis pipeline.

\subsection{RQ1: Detection Accuracy}
\label{sec:rq1}
We evaluate detection accuracy along two dimensions: \emph{effectiveness}---whether \tool detects known vulnerabilities---and \emph{precision}---whether the alerts it produces are true positives.
We count a true positive when the proof-of-concept exploit triggers an alert at a security-critical sink in its exploit chain, with a provenance DAG traceable to the taint source. A handful of cases alert at a different sink than the benchmark's nominal CWE label;
these are detections of real dangerous flows, and we note the CWE divergence where relevant.

\subsubsection{Detection Effectiveness}
We evaluate effectiveness on three benchmarks: SecBench.js, NodeMedic's benchmark, and a curated set of twelve real-world application vulnerabilities.

\noindent\textbf{SecBench.js:}
SecBench.js contains 503 CVEs across 4 CWE categories; 10 are not validated cases (package removed from npm, or core-API removal that crashes the library), leaving 493 in-scope cases against which we report the result in Table~\ref{tab:secbench-recall}.

\begin{table}[t]
\centering
\caption{
    Vulnerability detection accuracy on SecBench.js.
    \tool consistently outperforms the baseline NodeMedic-FINE.
}
\label{tab:secbench-recall}
\begin{tabular}{llrrrrr}
\toprule
&
& \multicolumn{1}{c}{\textbf{SecBench}} 
& \multicolumn{2}{c}{\textbf{\tool}} 
& \multicolumn{2}{c}{\textbf{NodeMedic-F}} \\
\cmidrule(lr){3-3} \cmidrule(lr){4-5} \cmidrule(lr){6-7}
\textbf{Category} & \textbf{CWE} & In-Scope & TP & Recall & TP & Recall \\
\midrule
Code Inj.     & 94   &  40 &  \underline{39} & \underline{97.5\%} & 19 & 47.5\% \\
Cmd Inj.      & 78   &  96 &  \underline{93} & \underline{96.9\%} & 68 & 70.8\% \\
Path Trav.    & 22   & 167 & 167 & 100.0\% & -- & -- \\
Proto.\ Poll. & 1321 & 190 & 172 & 90.5\% & -- & -- \\
\midrule
\textbf{Total}  & & 493 & \underline{471} & \underline{95.5\%} & 87 & 63.9\% \\
\bottomrule
\end{tabular}
\vspace{-10px}
\end{table}
\begin{table*}[t]
\centering
\caption{
End-to-end detection on real-world vulnerabilities (2024--2026) across multiple platforms. 
}
\label{tab:realworld}

\begin{tabular}{lllcclrc}
\toprule
\textbf{\#} & \textbf{Platform(s)} & \textbf{Application} & \textbf{CVE / GHSA} & \textbf{CWE}
            & \textbf{CWE Name} & \textbf{DAG Nodes} & \textbf{Det.} \\
\midrule
1  & Node.js    & FUXA v1.2.9             & CVE-2025-69983      & 94    & Code Injection       & 15     & Full \\
2  & Node.js    & FUXA v1.2.9             & CVE-2026-25895      & 22    & Path Traversal       & 37 & Full \\
3  & Node.js    & PsiTransfer             & GHSA-xphh-5v4r-r3rx & 22    & Path Traversal       & 258     & Full \\
4  & Node.js    & n8n v1.123              & CVE-2026-27493      & 94    & Code Injection       & 278    & Partial  \\
5  & Node.js    & Signal K plugin         & CVE-2026-23515      & 78    & OS Command Injection & 4      & Full \\
6  & Node.js    & NestJS devtools         & CVE-2025-54782      & 94    & Code Injection       & 3      & Full \\
7  & Node.js    & Flowise v3.0.7          & GHSA-jv9m-vf54-chjj & 22    & Path Traversal       & 9      & Full \\
8  & Node.js    & Flowise v3.0.7          & GHSA-j44m-5v8f-gc9c & 22    & Path Traversal       & 9      & Full \\
9  & Node.js    & @react-router/node      & CVE-2025-61686      & 22    & Path Traversal       & 53     & Full \\
10 & Node.js    & set-in v2.0.4           & CVE-2026-26021      & 1321  & Prototype Pollution  & 1      & Full \\
11 & Node.js    & deepHas v1.0.7          & CVE-2026-25047      & 1321  & Prototype Pollution  & 1      & Full \\
12 & Node.js    & locutus v2.0.38         & CVE-2026-25521      & 1321  & Prototype Pollution  & 13      & Full \\
13 & Chromium + SpiderMonkey     & n8n                     & CVE-2025-52478      & 79    & Cross-Site Scripting & 1      & Full \\
14 & Chromium + SpiderMonkey     & openwebui               & CVE-2025-65959      & 79    & Cross-Site Scripting & 1      & Full \\
15 & Chromium + SpiderMonkey     & Prometheus              & CVE-2026-40179      & 79    & Cross-Site Scripting & 95     & Full \\
16 & CPython    & Open WebUI 0.6.9        & CVE-2026-44565      & 22    & Path Traversal       & 3      & Full \\
17 & CPython    & Astrbot                 & CVE-2025-55449      & 94    & Code Injection       & 2      & Partial \\
18 & CPython    & crawl4ai                & CVE-2026-26216      & 94    & Code Injection       & 2      & Full \\
19 & CPython    & DB-GPT                  & CVE-2024-10835      & 89    & SQL Injection        & 11      & Full \\

\bottomrule
\end{tabular}

\vspace{2pt}
\footnotesize
\centering
Det.: \textbf{Full} = primary/expected sink reached; \textbf{Partial} = an intermediate sink fires  but the highest-severity sink is missed.
\end{table*}

\tool achieves 95.5\% overall recall.
The 22 missed cases share a few structural root causes.
Eighteen are prototype-pollution misses: 16 lose per-element taint when the library decomposes a tainted compound value into fresh per-key bindings before the prototype write, and 2 are \emph{selection flows} where the tainted key only navigates to \texttt{Object.prototype} while the written value is constant---which an explicit value-flow DTA does not flag.
The remaining four are injection misses: two command-injection cases that pass intermittently under async timing and two cases lose taint in an AST parser.

To compare against prior work on a larger scale, we run the dynamic taint-analysis stage of NodeMedic-FINE~\cite{nodemedicfine2025} on the same SecBench.js exploits.
Its taint engine declares sinks only for code injection and command injection, so we compare on the 136 cases in those two categories.
Drivers are translated to its instrumentation interface, reusing the identical installed packages.
NodeMedic-FINE's DTA detects 87 of 136 in-scope cases (63.9\%); \tool detects 132 of 136 (97.1\%) on the same categories.
NodeMedic-FINE uniquely detects only 1 case (timing-flaky under our engine), while \tool uniquely detects 46---28 where NodeMedic-FINE loses taint before the sink, and 18 where its Babel layer cannot instrument the package.

\noindent\textbf{NodeMedic dataset reproduction.}
To provide a direct comparison against prior work, we reproduce the 21-package evaluation dataset from NodeMedic~\cite{nodemedic2024}.
This dataset, curated from Ichnaea~\cite{karim2018platform} and Synode~\cite{staicu2018understanding}, contains 18 confirmed vulnerabilities and 3 true negatives.
Package \texttt{node-libnotify} has been removed from the npm registry so we evaluate the remaining 20.
\tool detects all 17 true positives and correctly produces no alert on all 3 true negatives.

\noindent\textbf{Real-world applications:}
We evaluate detection on 12 vulnerabilities disclosed in 2024--2026 across 10 production Node.js applications (Table~\ref{tab:realworld}), ranging from IoT plugins to enterprise platforms. Each case is instrumented with a single \texttt{\%SetTaint} call at the HTTP boundary; no application code is modified.
\tool fully detects 11 of the 12 vulnerabilities at their expected security-sensitive sinks. The remaining case (n8n) is a two-stage attack. We analyze the limitation in the case study below.

\begin{table}[t]
\centering
\caption{Precision evaluation on 141 patched packages.}
\label{tab:fp-analysis}
\begin{tabular}{lrrr}
\toprule
\textbf{CWE Category} & \textbf{Tests} & \textbf{True Neg.} & \textbf{False Pos.} \\
\midrule
Code Injection (CWE-94)      & 16 & 16 & 0 \\
Command Injection (CWE-78)   & 29 & 29 & 0 \\
Path Traversal (CWE-22)      & 9  & 9  & 0 \\
Prototype Pollution (CWE-1321) & 87 & 87 & 0 \\
\midrule
\textbf{Total}               & \textbf{141} & \textbf{141} & \textbf{0} \\
\bottomrule
\end{tabular}
\vspace{-10px}
\end{table}

\subsubsection{Precision}
For each SecBench.js case with an available patched version, we install the fixed package and rerun the identical exploit input; an alert on patched code is counted as a false positive.
Table~\ref{tab:fp-analysis} shows no alerts on 141 patched-version runs.
For vulnerable-version runs, we manually inspect each alert's sink and provenance DAG; all counted detections carry attacker taint to a security-sensitive operation in the exploit chain.
Some alerts reach a different dangerous sink than the benchmark's nominal CWE, which we record as CWE divergence rather than a false positive.

\subsubsection{Case Study: n8n}
\label{sec:casestudy-n8n}
n8n~\cite{n8n} is an 8{,}772-file TypeScript workflow platform built on Express, TypeORM, SQLite, and an expression engine.
Its vulnerability is a two-stage attack: an unauthenticated form payload is first saved to SQLite, then later read back and evaluated as code.
\tool tracks the first-stage flow from HTTP input to three \texttt{node\_sqlite3.Statement} database writes,
producing provenance DAGs up to 278 nodes using general propagation rules.
The run produces five alerts: three expected database-write alerts and two false filesystem alerts caused by over-propagation through TypeORM object merging, where tainted field values bleed onto metadata strings that later reach \texttt{existsSync}.
We classify the CVE as partially detected because \tool observes the write-side flow but cannot reconnect it to the later read path reaching \texttt{new Function()};
unlike file I/O, SQL persistence requires recovering the storage key space from queries, which would require SQL-aware bridging.

\begin{table}[t]
\centering
\setlength{\tabcolsep}{2.5pt}
\caption{Performance on NodeMedic's benchmark.}
\label{tab:nodemedic-overhead}
\begin{tabular}{@{}llrrr@{}}
\toprule
\textbf{Configuration} & \textbf{Node Version} &
\textbf{Base (ms)} & \textbf{Time (ms)} & \textbf{Overhead} \\
\midrule
Node, full JITs  & 24.13.1 & 68.5 & 68.5      & 1.00$\times$ \\
Node, Maglev     & 24.13.1 & 68.5 & 87.8      & 1.28$\times$ \\
\tool            & 24.13.1 & 68.5 & 126.4     & \textbf{1.85$\times$} \\
Jalangi          & 15.5.0 & 69.2 & 1{,}991.8 &
    28.78$\times$ \\
NodeMedic        & 15.5.0 & 69.2 & 3{,}516.6 &
    50.82$\times$ \\
NodeMedic-FINE   & 20.20.2 & 78.8 & 3{,}307.6 &
    41.95$\times$ \\
\bottomrule
\end{tabular}
\vspace{-10px}
\end{table}

\subsection{RQ2: Performance}
\label{sec:rq2}
\begin{table}[t]
\centering
\caption{Rule distribution by API category. Numbers denote action
  occurrences; P, C, and IE denote \texttt{propagate}, \texttt{clear}, and \texttt{inject}/\texttt{extract}, respectively.}
\label{tab:construct-coverage}
\setlength{\tabcolsep}{3.5pt}
\begin{tabular}{llrrrrr}
\toprule
\textbf{Category} & \textbf{Libraries} & \textbf{P} & \textbf{C} & \textbf{IE} & \textbf{Sink} &  \textbf{Set} \\
\midrule
String manip.   & String.prototype          & $49$ & $8$ & $4$  &    &   \\
Container/HOF   & Array, Map, Set           & $24$ &   & $34$ & $1$  &   \\
Encoding        & Buffer, URI, RegExp       & $23$ & $1$ &    &    &   \\
Async flow      & Promise, Generator        &  $2$ &   & $4$  &    &   \\
Object/JSON     & Object, JSON              &  $4$ &   &    & $1$  &   \\
Code injection  & eval, Function, vm        &      &   &    & $7$  &   \\
Command inj.    & child\_process            &      &   &    & $2$  &   \\
Filesystem      & fs                        &  $3$ &   &    & $15$ & $2$ \\
DOM/Browser     & Element, Document...      &  $9$ &   &    & $17$ & $1$ \\
Network/SQL     & http, sqlite3             &  $1$ &   &    & $2$ &   \\
Proto.\ poll.   & bytecode property store   &  $2$ &   &    & $10$ & $2$ \\
\bottomrule
\end{tabular}
\end{table}

We adopt NodeMedic's performance benchmark~\cite{nodemedic2024} which is the same as the reproduced benchmark in RQ1.
For each package, we reproduce NodeMedic's methodology: discover and freeze the public entry points, load the package, and invoke every entry point once with the same object supplied to each argument.
Each DTA configuration marks this object as tainted through its respective source interface.

Following NodeMedic's protocol, each configuration uses three warmups and ten fresh-process runs per package, interleaved with the baseline and timed externally.
Installation, entry-point discovery, and driver generation are excluded.
We first compute each package's mean runtime and overhead, then report the geometric mean across the 20 packages.

We directly measure six configurations on the same machine and frozen workload as shown in Table~\ref{tab:nodemedic-overhead}.
Source-rewriting tools run on their required Node versions, so we compare normalized slowdowns rather than absolute times.

Table~\ref{tab:nodemedic-overhead} reports 1.85$\times$ end-to-end overhead for \tool. Of this, Maglev-only execution accounts for 1.28$\times$, leaving 1.44$\times$ over the matched Maglev baseline for DTA.
On the same workload, source-rewriting systems are 22.7--27.5$\times$ slower on the identical workflow; the Jalangi-Babel instrumentation with no taint tracking already costs 15.6$\times$.
We report only our direct reruns because artifacts, dependencies, hardware,  and package availability differ from the original NodeMedic study.

\subsection{RQ3: Expressiveness and Automatic Synthesis}
\label{sec:rq3}

RQ3 evaluates the declarative rule layer itself.
We study whether its actions cover diverse API behaviors and whether new rules can be synthesized with minimal effort.

\noindent\textbf{Rule coverage.}
Table~\ref{tab:construct-coverage} summarizes \dsl action usage across 11 API categories in the final 303-LoC specification, covering both Node.js and Chromium contexts.
Library summaries rely mainly on \texttt{propagate} and \texttt{inject}/\texttt{extract}; detection rules use guarded \texttt{sink} actions; and \texttt{set} models cross-boundary state such as file I/O.
New detection surfaces required rule changes only: CWE-89 support added one \texttt{sink} rule for \texttt{node\_sqlite3.Statement}, while DOM support added DOM rules without changing the Shadow VM core.

\begin{table}[t]
\centering
\caption{Node.js API coverage compared with CodeQL.}
\label{tab:api-coverage-ql}
\begin{tabular}{lrrrr}
\toprule
\textbf{Language} & \textbf{APIs} & \textbf{Shared} & \textbf{Missing} & \textbf{LoC} \\
\midrule
CodeQL~\cite{codeql} & 139 & 135 & 8 & 1,149 \\
\dsl                 & 143 & 135 & 4 & \textbf{63} \\
\bottomrule
\end{tabular}
\vspace{-10px}
\end{table}

\noindent\textbf{Coverage comparison with CodeQL.}
We also compare LLM-synthesized \dsl rules for Node.js standard-library against CodeQL's taint specification rules (Table~\ref{tab:api-coverage-ql}). 
Our rules cover $97.2\%$ of the APIs modeled in CodeQL with $18\times$ fewer LoC, indicating that \dsl is expressive enough while being a lot more concise.
Further, our rules have source, sink, and summary models $97\%$ match that provided by CodeQL, underpinning AI-friendliness of \dsl.
The few discrepancies reflect a difference in analysis style rather than coverage: where CodeQL abstracts an operation as a sink or source (e.g., treating every file read as tainted), \dsl tracks the actual data through a precise I/O bridge, trading CodeQL's soundness-oriented over-approximation for runtime precision.

\noindent\textbf{LLM-assisted rule authoring.}
What if LLM synthesizes buggy \dsl rules due to unfamiliarity?
The synthesis undergoes a validation process:
when we task LLMs to synthesize \dsl rules, the model is asked to simultaneously generate smoke tests, which will be executed alongside \tool and the rule.
After execution, it receives one optional missing-model diagnostic containing the observed host event and arity, but not the rule or taint locations.
As such, feedbacks from the validator can be used to help LLMs refine their generation.

\begin{table}[t]
\centering
\caption{LLM-assisted rule authoring by Claude Opus 4.8 with one diagnostic feedback step.}
\label{tab:llm-authoring}
\begin{tabular}{lrr}
\toprule
\textbf{Metric} & \textbf{Initial} & \textbf{After feedback} \\
\midrule
Public APIs covered & $56$ & $56$ {\color{gray}($\pm$0)} \\
Compiled generated rules & $68$ & $72$ {\color{CodeGreen}($+$4)} \\ \midrule
Compiling candidate patterns & $32/32$ & $32/32$ {\color{gray}($\pm$0)} \\ 
Provided action-class realization & $30/32$ & $30/32$ {\color{gray}($\pm$0)} \\
Runtime candidate behaviors passed & $19/32$ & $29/32$ {\color{CodeGreen}($+$10)} \\
Raw smoke-test checks passed & $41/54$ & $51/54$ {\color{CodeGreen}($+$10)} \\
\bottomrule
\end{tabular}
\end{table}

Table~\ref{tab:llm-authoring} shows a partial result on 32 taint-behavior patterns covering 56 public APIs from \texttt{Buffer}, \texttt{zlib}, \texttt{URLSearchParams}, and \texttt{Array}.
Initial synthesis is already compiling, and
one round of feedback produces a gain of ten passing behaviors ($19/32$ to $29/32$), mainly by binding rules to lower-level host events.
The three remaining failures involve callback-return association in Array higher-order functions.

\subsection{RQ4: Portability}
\label{sec:rq4}

\begin{table}[t]
\centering
\caption{
  Statistics of lines-of-code (LoC) in the engine implementation language (C++) across three runtime instantiations for \tool adaptation.
  ``---'' indicates not applicable.}
\label{tab:impl-scale}
\setlength{\tabcolsep}{4pt}
\begin{tabular}{lrrr}
\toprule
\textbf{Component} & \textbf{V8} & \textbf{CPython} & \textbf{SpiderMonkey} \\
\midrule
\multicolumn{4}{l}{\emph{Shared (unchanged across runtimes)}} \\
\quad Taint engine + graph    & 1,004 & 1,004 & 1,004 \\
\quad \dsl interpreter         & 1,494 & 1,494 & 1,494 \\
\quad Rule compiler           &   433 &   433 &   433 \\
\quad Logger                  & 1,010 & 1,010 & 1,010 \\
\midrule
\multicolumn{4}{l}{\emph{Per-runtime adapter}} \\
\quad Adapter layer           & 2,181 &   446 & 661 \\
\quad Interpreter hooks       & 1,685 &   528 & 236 \\
\quad JIT integration         & 1,278 &   --- & --- \\
\quad Runtime stubs           &   530 &   98 & 225 \\
\quad Host state + IPC + GC   &   497 &   20 & 64 \\
\midrule
\textbf{Shared core}         & \textbf{3,941} & \textbf{3,941} & \textbf{3,941} \\
\textbf{Engine-specific}     & \textbf{6,171} & \textbf{1,092} & \textbf{1,186} \\
\midrule
\dsl                         & 303    & 235   & 138 \\
\bottomrule
\end{tabular}
\vspace{-10px}
\end{table}

The Shadow VM is built against a general host interface (\S\ref{sec:3.3}), and a new runtime needs only (i) an adapter that satisfies the interface and (ii) runtime-specific rules. We test this along three axes: \emph{embedding} (same engine, different host), \emph{engine} (same language, different engine), and \emph{language} (different language). For each axis we report porting cost in Table~\ref{tab:impl-scale} and CVE detection in Table~\ref{tab:realworld}.

\noindent\textbf{Embedding: V8 in Chromium.}
The same V8 implementation runs unmodified in Chromium's renderer process.
The only additions are 27 DOM rules and the host-operation bindings.
The Shadow VM core and all propagation rules are unchanged.
The instantiation passes 21/22 DOM XSS test cases and detects three client-side XSS CVEs (Table~\ref{tab:realworld}).
The single failing test is a serialization boundary: Blink's \texttt{SerializedScriptValue} path for \texttt{postMessage} bypasses V8's serializer hooks, so taint is dropped as data crosses into Blink.

\noindent\textbf{Engine: SpiderMonkey.}
The SpiderMonkey port reuses the shared core and adds 1{,}186 engine-specific LoC.
Rule reuse is the main result: the complete V8 rule file compiles unchanged on SpiderMonkey's toolchain, and every rule \emph{body} ports byte-for-byte.
Only the operation keys are rebound to Spidermonkey builtins or intrinsics. SpiderMonkey implements as self-hosted JavaScript, the key must target SpiderMonkey's intrinsic name rather than the V8 builtin name. With rebinding, 18 of 19 ported behaviors fire; the failure is \texttt{concat}, whose SpiderMonkey implementation does not expose a call boundary and therefore needs a fixed opcode transition. 
The same three client-side CVEs are also detected on SpiderMonkey. 

\noindent\textbf{Language: CPython.}
Porting to CPython required a 446-LoC adapter and 528 LoC of bytecode instrumentation hooks in \texttt{ceval.c}.
LLM authored 138-LoC \dsl rules and pass 222 tests. 
End to end, the CPython instantiation fully detects three real-world CVEs and partially detects one (Table~\ref{tab:realworld}). 
The partial case, Astrbot \cite{astrbotcve}, is a two-stage attack whose taint is lost at the file-write boundary because \texttt{f.write} does not expose the destination path attached to \texttt{f} to the rule; bridging it would require querying additional object metadata in \dsl.

Table~\ref{tab:impl-scale} shows the shared core (3{,}941~LoC) is identical across all three runtimes; porting effort is concentrated in adapters, operation bindings, and rules.
These per-runtime figures are not directly comparable: V8 includes JIT integration and Chromium's IPC, whereas the CPython and SpiderMonkey ports cover the interpreter path only.
The remaining misses across axes are unbridged serialization or persistence boundaries rather than failures to reuse taint semantics.

\section{Discussion}
\label{sec:discussion}

\noindent\textbf{Scope.} \tool focuses on six CWEs. Additional vulnerability classes require new rules to describe the oracle.
\tool, as related analyses~\cite{karim2018platform,nodemedic2024,staicu2018understanding} does not consider implicit flows.
Implicit flow tracking produces no demonstrated detection improvement for this vulnerability class~\cite{calzavara2025,staicu2019empirical} while incurring substantial overhead (36.7$\times$ in ~\cite{panoptichrome2024}). A future extension is a condition mark action that marks tainted branch conditions in provenance, combined with concolic exploration to determine sink reachability along controlled paths.

Like related tools~\cite{karim2018platform,nodemedic2024,staicu2018understanding}, \tool relies on rules by precise analysis of native operations. This is a shared limitation of all instrumentation-based dynamic analyses~\cite{schwartz2010dta}. Future work includes LLM-assisted rule generation for unmodeled native APIs and selective instrumentation that activates hooks only on code reachable from taint sources to reduce overhead.

\noindent\textbf{Engineering limitations.}
Our performance evaluation measures short, end-to-end package-analysis executions and does not characterize sustained CPU-intensive workloads.
Always-on hooks for frequent operations, particularly property accesses and function calls, can impose higher overhead on computation-heavy applications.
We leave selective instrumentation and full Turbofan-tier JIT integration for future work.

Whole-payload tainting may conflate tainted and untainted fields through object spread or \texttt{Object.assign}; field-sensitive source tainting eliminates this without engine modification.
The dominant false-negative pattern is string decomposition: operations like \texttt{String.split} produce new values that lose association with the original tainted input.
This is shared by all explicit-flow DTA tools; character-level tracking~\cite{Foxhound} addresses it at engine-specific cost.
Taint does not survive external storage boundaries without explicit bridging rules, which we leave to future work.

\noindent\textbf{Threats to validity.}
SecBench.js contains CVEs from 2017--2022; many are small single-file packages that may not represent modern application complexity. 
We mitigate this with the NodeMedic dataset~\cite{nodemedic2024} and 19 manually collected vulnerabilities from 2024--2026 in larger-scale production applications up to 8{,}772 files. 

\section{Related Works}

\subsection{Dynamic Taint Analysis}

Dynamic taint analysis has been formalized by Schwartz et al.~\cite{schwartz2010dta} and instantiated across execution environments: binary-level~\cite{taintcheck2005,libdft2012,dytan2007,selectivetaint2021}, managed runtimes~\cite{taintdroid2014,dtaplusplus2011}, and
JavaScript engines. All uniformly hardcode taint semantics into their instrumentation layer.
For JavaScript, source-rewriting tools like Jalangi2~\cite{jalangi2013} and its extensions~\cite{nodemedic2024,nodemedicfine2025} are engine-agnostic, but require analyzed modules to pass a source-transformation pipeline and model native builtins externally; NodeMedic adds Babel to broaden syntax support. Engine-native tools embed taint tracking directly: Foxhound~\cite{Foxhound} modifies SpiderMonkey's string representation; PanoptiChrome~\cite{panoptichrome2024} instruments V8's Ignition interpreter with implicit flow tracking. Client-side analyses~\cite{JAW,melicher2018riding,probetheproto2022,kang2025follow} target DOM XSS and prototype pollution in browsers.
\tool differs from prior DTA tools for stock JavaScript engines in separating taint \emph{semantics}
(what to track) from \emph{mechanism} (how to track): the Shadow VM provides
observation infrastructure while \dsl rules specify propagation behavior.

\subsection{Function Summaries and Taint Specification Languages}
Interprocedural taint analysis relies on function summaries that abstract callees into input-to-output transfer functions~\cite{reps1995ifds,sharir1981twoApproach}. In static analysis, FlowDroid~\cite{flowdroid2014} precomputes summaries for Android framework methods; DroidSafe~\cite{droidsafe2015} and JN-SAF~\cite{jnsaf2018} extend summaries across native boundaries; CompTaint~\cite{comptaint2023} demonstrates industrial scaling at AWS via cached API models. DSLs externalize summary authoring: CodeQL~\cite{codeql,avgustinov2016ql} encodes taint flow as Datalog queries, PQL~\cite{pql2005} matches taint patterns across object histories, and fluentTQL~\cite{fluenttql,secucheck2022} embeds taint queries as typed Java method chains. Recent work has also begun to synthesize such taint specifications for CodeQL with LLMs~\cite{li2024iris,wang2025qlcoder}, mirroring our language-model-friendly rule authoring. These approaches resolve specifications at analysis time via static solvers, not at program runtime.

For dynamic execution, Ichnaea~\cite{karim2018platform} is closest: it replays taint on a Jalangi-level abstract machine and models native functions with imperative hooks.
TruffleTaint~\cite{truffletaint}, Augur~\cite{aldrich2022augur}, and ALDA~\cite{cheng2022creating} also externalize analysis logic, but as programmatic policies tied to their host framework;
Whamm~\cite{whamm2025} shows that probes can be compiled into engine hooks, but provides no taint-specific semantics. 
\tool differs by combining a stock-runtime Shadow VM with a compiled taint-specific language: \dsl rules are executed by the Shadow VM, include formal \texttt{inject}/\texttt{extract} semantics for native/callback transfer, and reuse rule bodies across V8, SpiderMonkey, and CPython after operation rebinding.

\section{Conclusion}

We present \dsl, a declarative taint specification language, and \tool, an extensible, performant, and accurate Node.js dynamic taint analysis framework that executes \dsl over a runtime-independent Shadow Virtual Machine, which could be ported across Chromium, Spidermonkey and CPython.
In future work, we plan to improve precision across decomposition and persistence boundaries and to automate rule synthesis for unmodeled native APIs.

\section*{Acknowledgment}
This research is in part based upon work supported by credits supported through Amazon Nova AI Challenge: Trusted Software Agents.
Any opinions, findings, and conclusions or recommendations expressed in this material are those of the author(s) and do not necessarily reflect the views of the sponsors.
Language model assistants were used to help polish the paper.

\bibliographystyle{IEEEtran}
\bibliography{reference}

@inproceedings{jalangi2013,
  author    = {Koushik Sen and Swaroop Kalasapur and Tasneem G. Brutch and Simon Gibbs},
  title     = {Jalangi: A Selective Record-Replay and Dynamic Analysis Framework for {JavaScript}},
  booktitle = {ESEC/FSE},
  year      = {2013},
  pages     = {488--498},
  doi       = {10.1145/2491411.2491447},
}

@inproceedings{nodemedic2024,
  author    = {Darion Cassel and Wai Tuck Wong and Limin Jia},
  title     = {{NodeMedic}: End-to-End Analysis of {Node.js} Vulnerabilities with Provenance Graphs},
  booktitle = {2023 IEEE 8th European Symposium on Security and Privacy (EuroS{\&}P)},
  year      = {2023},
  pages     = {1101-1127},
  doi       = {10.1109/EuroSP57164.2023.00068},
}

@inproceedings{Foxhound,
  author = {David Klein and Thomas Barber and Souphiane Bensalim and Ben Stock and Martin Johns},
  title = {Hand Sanitizers in the Wild: A Large-scale Study of Custom JavaScript Sanitizer Functions},
  booktitle = {Proc. of the IEEE European Symposium on Security and Privacy},
  year = {2022},
  month = jun,
}

@inproceedings{nodemedicfine2025,
  author    = {Darion Cassel and Nuno Sabino and Min-Chien Hsu and Ruben Martins and Limin Jia},
  title     = {{NodeMedic-FINE}: Automatic Detection and Exploit Synthesis for {Node.js} Vulnerabilities},
  booktitle = {NDSS},
  year      = {2025},
  doi       = {10.14722/ndss.2025.241636},
}

@inproceedings{panoptichrome2024,
  author    = {Rahul Kanyal and Smruti Ranjan Sarangi},
  title     = {{PanoptiChrome}: A Modern In-browser Taint Analysis Framework},
  booktitle = {WWW},
  year      = {2024},
  doi       = {10.1145/3589334.3645699},
}

@inproceedings{probetheproto2022,
  author    = {Zifeng Kang and Song Li and Yinzhi Cao},
  title     = {Probe the Proto: Measuring Client-Side Prototype Pollution Vulnerabilities of One Million Real-world Websites},
  booktitle = {NDSS},
  year      = {2022},
}

@inproceedings{domsday2018,
  author    = {William Melicher and Anupam Das and Mahmood Sharif and Lujo Bauer and Limin Jia},
  title     = {Riding out {DOMsday}: Towards Detecting and Preventing {DOM} Cross-Site Scripting},
  booktitle = {NDSS},
  year      = {2018},
}

@inproceedings{JAW,
  title = {JAW: Studying Client-side CSRF with Hybrid Property Graphs and Declarative Traversals},
  author= {Soheil Khodayari and Giancarlo Pellegrino},
  booktitle = {30th {USENIX} Security Symposium ({USENIX} Security 21)},
  year = {2021},
  address = {Vancouver, B.C.},
  publisher = {{USENIX} Association},
}

@inproceedings{melicher2018riding,
  title={Riding out domsday: Towards detecting and preventing dom cross-site scripting},
  author={Melicher, William and Das, Anupam and Sharif, Mahmood and Bauer, Lujo and Jia, Limin},
  booktitle={2018 Network and Distributed System Security Symposium (NDSS)},
  year={2018}
}

@inproceedings{staicu2018understanding,
  title={Understanding and automatically preventing injection attacks on Node. js},
  author={Staicu, Cristian-Alexandru and Pradel, Michael and Livshits, Ben},
  booktitle={Network and Distributed System Security Symposium (NDSS)},
  year={2018}
}

@inproceedings{calzavara2025,
  author    = {Stefano Calzavara and Samuele Casarin and Riccardo Focardi},
  title     = {Dynamic Security Analysis of {JavaScript}: Are We There Yet?},
  booktitle = {WWW},
  year      = {2025},
  pages     = {1105--1115},
  doi       = {10.1145/3696410.3714614},
}

@article{explodejs2025,
  author    = {Filipe Marques and Mafalda Ferreira and Andr{\'e} Nascimento and Miguel E. Coimbra and Nuno Santos and Limin Jia and Jos{\'e} Fragoso Santos},
  title     = {Automated Exploit Generation for {Node.js} Packages},
  journal   = {Proc. ACM Program. Lang.},
  volume    = {9},
  number    = {PLDI},
  pages     = {1341--1366},
  year      = {2025},
  doi       = {10.1145/3729304},
}

@misc{codeql,
  author    = {{GitHub}},
  title     = {{CodeQL}: Semantic Code Analysis Engine},
  year      = {2024},
  howpublished = {\url{https://codeql.github.com/}},
}

@INPROCEEDINGS{secbenchjs,
  author={Bhuiyan, Masudul Hasan Masud and Parthasarathy, Adithya Srinivas and Vasilakis, Nikos and Pradel, Michael and Staicu, Cristian-Alexandru},
  booktitle={2023 IEEE/ACM 45th International Conference on Software Engineering (ICSE)}, 
  title={SecBench.js: An Executable Security Benchmark Suite for Server-Side JavaScript}, 
  year={2023},
  pages={1059-1070},
  doi={10.1109/ICSE48619.2023.00096},
}

@misc{n8n,
  title     = {n8n - Secure Workflow Automation for Technical Teams},
  year      = {2026},
  howpublished = {\url{https://github.com/n8n-io/n8n/}},
  note         = {Accessed: 2026-05-17}
}

@misc{reactroutercve,
title = {CVE-2025-61686},
year = {2026},
howpublished = {\url{https://nvd.nist.gov/vuln/detail/CVE-2025-61686}},
note = {Accessed: 2026-05-17}
}

@misc{astrbotcve,
title = {CVE-2025-55449},
year = {2026},
howpublished = {\url{https://nvd.nist.gov/vuln/detail/CVE-2025-55449}},
note = {Accessed: 2026-05-17}
}

@misc{reactrouterdownload,
title = {React Router},
year = {2026},
howpublished = {\url{https://reactrouter.com/}},
note = {Accessed: 2026-05-17}
}

@inproceedings{reps1995ifds,
  title={Precise interprocedural dataflow analysis via graph reachability},
  author={Reps, Thomas and Horwitz, Susan and Sagiv, Mooly},
  booktitle={Proceedings of the 22nd ACM SIGPLAN-SIGACT symposium on Principles of programming languages},
  pages={49--61},
  year={1995}
}

@article{sharir1981twoApproach,
  title={Two approaches to interprocedural data flow analysis},
  author={Pnueli, M and Sharir, Micha},
  journal={Program flow analysis: theory and applications},
  pages={189--234},
  year={1981}
}

@article{flowdroid2014,
  title={Flowdroid: Precise context, flow, field, object-sensitive and lifecycle-aware taint analysis for android apps},
  author={Arzt, Steven and Rasthofer, Siegfried and Fritz, Christian and Bodden, Eric and Bartel, Alexandre and Klein, Jacques and Le Traon, Yves and Octeau, Damien and McDaniel, Patrick},
  journal={ACM sigplan notices},
  volume={49},
  number={6},
  pages={259--269},
  year={2014},
  publisher={ACM New York, NY, USA}
}

@inproceedings{droidsafe2015,
  title={Information flow analysis of android applications in droidsafe.},
  author={Gordon, Michael I and Kim, Deokhwan and Perkins, Jeff H and Gilham, Limei and Nguyen, Nguyen and Rinard, Martin C},
  booktitle={NDSS},
  volume={15},
  number={201},
  pages={110},
  year={2015}
}

@inproceedings{jnsaf2018,
  title={Jn-saf: Precise and efficient ndk/jni-aware inter-language static analysis framework for security vetting of android applications with native code},
  author={Wei, Fengguo and Lin, Xingwei and Ou, Xinming and Chen, Ting and Zhang, Xiaosong},
  booktitle={Proceedings of the 2018 ACM SIGSAC Conference on Computer and Communications Security},
  pages={1137--1150},
  year={2018}
}

@inproceedings{comptaint2023,
  title={Compositional taint analysis for enforcing security policies at scale},
  author={Banerjee, Subarno and Cui, Siwei and Emmi, Michael and Filieri, Antonio and Hadarean, Liana and Li, Peixuan and Luo, Linghui and Piskachev, Goran and Rosner, Nicol{\'a}s and Sengupta, Aritra and others},
  booktitle={Proceedings of the 31st ACM Joint European Software Engineering Conference and Symposium on the Foundations of Software Engineering},
  pages={1985--1996},
  year={2023}
}

@inproceedings{avgustinov2016ql,
  title={QL: Object-oriented queries on relational data},
  author={Avgustinov, Pavel and De Moor, Oege and Jones, Michael Peyton and Sch{\"a}fer, Max},
  booktitle={30th European Conference on Object-Oriented Programming (ECOOP 2016)},
  pages={2--1},
  year={2016},
  organization={Schloss Dagstuhl--Leibniz-Zentrum f{\"u}r Informatik}
}

@article{pql2005,
  title={Finding application errors and security flaws using PQL: a program query language},
  author={Martin, Michael and Livshits, Benjamin and Lam, Monica S},
  journal={Acm Sigplan Notices},
  volume={40},
  number={10},
  pages={365--383},
  year={2005},
  publisher={ACM New York, NY, USA}
}

@article{fluenttql,
  title={Fluently specifying taint-flow queries with fluent tql},
  author={Piskachev, Goran and Sp{\"a}th, Johannes and Budde, Ingo and Bodden, Eric},
  journal={Empirical Software Engineering},
  volume={27},
  number={5},
  pages={104},
  year={2022},
  publisher={Springer}
}

@inproceedings{secucheck2022,
  title={Secucheck: Engineering configurable taint analysis for software developers},
  author={Piskachev, Goran and Krishnamurthy, Ranjith and Bodden, Eric},
  booktitle={2021 IEEE 21st International Working Conference on Source Code Analysis and Manipulation (SCAM)},
  pages={24--29},
  year={2021},
  organization={IEEE}
}

@inproceedings{selectivetaint2021,
  title={$\{$SelectiveTaint$\}$: Efficient data flow tracking with static binary rewriting},
  author={Chen, Sanchuan and Lin, Zhiqiang and Zhang, Yinqian},
  booktitle={30th USENIX Security Symposium (USENIX Security 21)},
  pages={1665--1682},
  year={2021}
}

@article{karim2018platform,
  title={Platform-independent dynamic taint analysis for javascript},
  author={Karim, Rezwana and Tip, Frank and Sochurkova, Alena and Sen, Koushik},
  journal={IEEE Transactions on Software Engineering},
  volume={46},
  number={12},
  pages={1364--1379},
  year={2018},
  publisher={IEEE}
}

@article{whamm2025,
  title={Debugging webassembly? put some whamm on it!},
  author={Gilbert, Elizabeth and Schneider, Matthew and An, Zixi and Thalanki, Suhas and Bowman, Wavid and Bai, Alexander Y and Titzer, Ben L and Miller, Heather},
  journal={Proceedings of the ACM on Programming Languages},
  volume={9},
  number={OOPSLA2},
  pages={2058--2086},
  year={2025},
  publisher={ACM New York, NY, USA}
}

@inproceedings{schwartz2010dta,
  title={All you ever wanted to know about dynamic taint analysis and forward symbolic execution (but might have been afraid to ask)},
  author={Schwartz, Edward J and Avgerinos, Thanassis and Brumley, David},
  booktitle={2010 IEEE symposium on Security and privacy},
  pages={317--331},
  year={2010},
  organization={IEEE}
}

@inproceedings{taintcheck2005,
  title={Dynamic taint analysis for automatic detection, analysis, and signaturegeneration of exploits on commodity software.},
  author={Newsome, James and Song, Dawn Xiaodong and others},
  booktitle={NDSS},
  volume={5},
  pages={3--4},
  year={2005}
}

@inproceedings{libdft2012,
  title={libdft: Practical dynamic data flow tracking for commodity systems},
  author={Kemerlis, Vasileios P and Portokalidis, Georgios and Jee, Kangkook and Keromytis, Angelos D},
  booktitle={Proceedings of the 8th ACM SIGPLAN/SIGOPS conference on Virtual Execution Environments},
  pages={121--132},
  year={2012}
}

@inproceedings{dytan2007,
  title={Dytan: a generic dynamic taint analysis framework},
  author={Clause, James and Li, Wanchun and Orso, Alessandro},
  booktitle={Proceedings of the 2007 international symposium on Software testing and analysis},
  pages={196--206},
  year={2007}
}

@article{taintdroid2014,
  title={Taintdroid: an information-flow tracking system for realtime privacy monitoring on smartphones},
  author={Enck, William and Gilbert, Peter and Han, Seungyeop and Tendulkar, Vasant and Chun, Byung-Gon and Cox, Landon P and Jung, Jaeyeon and McDaniel, Patrick and Sheth, Anmol N},
  journal={ACM Transactions on Computer Systems (TOCS)},
  volume={32},
  number={2},
  pages={1--29},
  year={2014},
  publisher={ACM New York, NY, USA}
}

@inproceedings{dtaplusplus2011,
  title={Dta++: dynamic taint analysis with targeted control-flow propagation.},
  author={Kang, Min Gyung and McCamant, Stephen and Poosankam, Pongsin and Song, Dawn and others},
  booktitle={Ndss},
  year={2011}
}

@inproceedings{kang2025follow,
  title={Follow my flow: Unveiling client-side prototype pollution gadgets from one million real-world websites},
  author={Kang, Zifeng and Lyu, Muxi and Liu, Zhengyu and Yu, Jianjia and Fan, Runqi and Li, Song and Cao, Yinzhi},
  booktitle={2025 IEEE Symposium on Security and Privacy (SP)},
  pages={991--1008},
  year={2025},
  organization={IEEE}
}

@inproceedings{staicu2019empirical,
  title={An empirical study of information flows in real-world javascript},
  author={Staicu, Cristian-Alexandru and Schoepe, Daniel and Balliu, Musard and Pradel, Michael and Sabelfeld, Andrei},
  booktitle={Proceedings of the 14th ACM SIGSAC Workshop on Programming Languages and Analysis for Security},
  pages={45--59},
  year={2019}
}

@inproceedings{truffletaint,
author = {Kreindl, Jacob and Bonetta, Daniele and Stadler, Lukas and Leopoldseder, David and M\"{o}ssenb\"{o}ck, Hanspeter},
title = {Multi-language dynamic taint analysis in a polyglot virtual machine},
year = {2020},
isbn = {9781450388535},
publisher = {Association for Computing Machinery},
address = {New York, NY, USA},
booktitle = {Proceedings of the 17th International Conference on Managed Programming Languages and Runtimes},
pages = {15–29},
numpages = {15},
location = {Virtual, UK},
series = {MPLR '20}
}

@inproceedings{aldrich2022augur,
  title={Augur: Dynamic taint analysis for asynchronous javascript},
  author={Aldrich, Mark W and Turcotte, Alexi and Blanco, Matthew and Tip, Frank},
  booktitle={Proceedings of the 37th IEEE/ACM International Conference on Automated Software Engineering},
  pages={1--4},
  year={2022}
}

@inproceedings{cheng2022creating,
  title={Creating concise and efficient dynamic analyses with ALDA},
  author={Cheng, Xiang and Devecsery, David},
  booktitle={Proceedings of the 27th ACM International Conference on Architectural Support for Programming Languages and Operating Systems},
  pages={740--752},
  year={2022}
}

@article{zhu2025locus,
  title={Locus: Agentic predicate synthesis for directed fuzzing},
  author={Zhu, Jie and Shen, Chihao and Li, Ziyang and Yu, Jiahao and Chen, Yizheng and Pei, Kexin},
  journal={arXiv preprint arXiv:2508.21302},
  year={2025}
}

@article{wang2025contemporary,
  title={A contemporary survey of large language model assisted program analysis},
  author={Wang, Jiayimei and Ni, Tao and Lee, Wei-Bin and Zhao, Qingchuan},
  journal={arXiv preprint arXiv:2502.18474},
  year={2025}
}

@article{simsek2025pocgen,
  title={Pocgen: Generating proof-of-concept exploits for vulnerabilities in npm packages},
  author={Simsek, Deniz and Eghbali, Aryaz and Pradel, Michael},
  journal={arXiv preprint arXiv:2506.04962},
  year={2025}
}

@article{li2024iris,
  title={{IRIS}: {LLM}-Assisted Static Analysis for Detecting Security Vulnerabilities},
  author={Li, Ziyang and Dutta, Saikat and Naik, Mayur},
  journal={arXiv preprint arXiv:2405.17238},
  year={2024}
}

@article{wang2025qlcoder,
  title={{QLCoder}: A Query Synthesizer for Static Analysis of Security Vulnerabilities},
  author={Wang, Claire and Li, Ziyang and Dutta, Saikat and Naik, Mayur},
  journal={arXiv preprint arXiv:2511.08462},
  year={2025}
}

\end{document}